\documentclass[11pt]{article}
\usepackage{colacl}

\title{Dictionaries merger for text expansion in question answering}
\author
{Bernard JACQUEMIN\\
\begin{tabular}{cc}
                                              & \\
ILPGA, Universit\'{e} de la Sorbonne Nouvelle & Xerox Research Centre Europe\\
19, rue des Bernardins                        & 6, chemin de Maupertuis\\
75\,005 Paris (France)                        & 38\,240 Meylan (France)\\
                                              &
\end{tabular}\\
Bernard.Jacquemin@xrce.xerox.com
}

\begin{document}
\maketitle

\begin{abstract}

This paper presents an original way to add new data in a reference
dictionary from several other lexical resources, without loosing
any consistence. This operation is carried in order to get lexical
information classified by the sense of the entry. This
classification makes it possible to enrich utterances (in QA: the
queries) following the meaning, and to reduce noise. An analysis
of the experienced problems shows the interest of this method, and
insists on the points that have to be tackled.

\end{abstract}

\section{Introduction}


Our society is currently facing an increasing amount of textual
data, that no-one can store up or even read. Many automatic
systems are designed to find a requested piece of information. All
the current systems use dictionaries to identify data in texts or
in queries. QA softwares, which are particularly demanding about
data from the dictionary, have a similar mode of working: they
process an utterance (generally the query) in order to provide the
largest number of way to express the same meaning. Then they try
to find a match between the expanded utterance and a text. For
example, \cite{Hull99} expands synonymically the 'significant'
vocabulary of the question. QUALC \cite{FerretAl99} adds stemming
expansion prior to using a search engine. The Falcon system
\cite{MoldovanAl00} uses some semantic relations from
\textit{WordNet} when it expands the question.


In this paper, I present a way to process dictionaries to make
them consistent with the needs of the application. I first
describe the lexical needs of my QA application. I secondly
outline the issue of the use of several incompatible dictionaries.
Then I show the way I distribute information from additional
dictionaries to a reference one: synonyms, derivative forms and
taxonomy. Finally, I present the problems and difficulties I
found.

\section{What the QA method needs}

The QA system \cite{Jacquemin03} is based on a matching procedure
between query and text segment. As most of the other approaches,
my methodology solves the problem of the different ways to express
the same idea by adding to the utterance ('enrichments') synonyms,
derivatives or words belonging to the same taxonomy.

My method entails two new features: First, it uses semantic
disambiguation in order to choose the right meaning to each word
in the sentences. I notice that most of the QA systems try to give
as many enrichments as possible to a word rather than to a
meaning. The answers often correspond to a sense different from
the original one. But if each enrichment has the same sense as the
original one, the noise decreases.



The fact that a semantic disambiguator needs a large context to
the word to be disambiguated \cite{Weaver49} provides the second
feature: the query generally comprises few words. I decided to
process the documents to build an enriched informative structure
\cite{Jacquemin04}. But this feature falls outside the scope of
this paper.

My semantic disambiguator \cite{JacqueminAl02} is an evolution of
a tool previously developed for both French and English at XRCE
\cite{Brun2000,BrunAl01}. The idea is to use a dictionary as a
tagged corpus to extract semantic disambiguation rules. The
contextual data (syntactic, lexico-syntactic and
semantico-syntactic) for a given sense of a word are seen as
differential indications. So when the schema is found in the
context of this word in a sentence, the corresponding sense is
assigned.

\begin{figure}[htb]
\begin{center}\small{
\mbox{\parbox{7cm}{
\textsf{Example from Dubois' dictionary (entry: \textit{remporter}):}\\
On \textbf{remporte} la victoire sur ses adversaires (sense nb 2 : \textit{gagner})\\
\textsl{We win a victory over our adversaries.}\\

\textsf{Dependency containing \textit{remporter}:}\\
\texttt{VARG[DIR](remporter,victoire)}\\

\textsf{Corresponding disambiguation rule:}\\
\texttt{remporter:\,VARG[DIR](remporter,victoire) ==> sense \textit{gagner}}\\
}}}
\caption{Extraction of a semantic disambiguation rule.}
\label{exDesamb}
\end{center}
\end{figure}

In figure \ref{exDesamb}, we can see how a disambiguation rule is
extracted from an informative field of Dubois' French dictionary
\cite{dubois-duboischarlier97}. From the instance field of the
entry \textit{remporter} in its second sense \textit{gagner} (to
win), the XIP parser \cite{Ait-al02} extracts a lexico-syntactic
schema: \texttt{VARG[DIR]} means that the argument
\textit{victoire} is a direct object of the argument
\textit{remporter}. The rule built from this dependency indicates
that the sense of the word \textit{remporter}, in a context where
\textit{victoire} (victory) is the direct object, is the second
sense \textit{gagner} (to win). Two other types of rules exists:
the first type puts lexical rules into general use, replacing
lexical arguments by corresponding semantic classes. The other one
uses syntactic schemas stipulated by the dictionary (for instance:
transitive, reflexive, etc.).


The dictionary needs of both QA system and semantic disambiguator
are of two natures. First, the dictionary is required to share out
data following sense and not following lemma: The data are
differential indications. Second, the dictionary is required to
contain contextual information as much as possible: examples or
collocations (lexical rules), semantic classes or application
domains (generalized rules), subcategorisation\dots The Dubois'
dictionary yields to these demands, and moreover it contains some
data that could be helpful to enrich an utterance: synonyms,
instructions for derivations\dots


\section{Enrichment problems}


Several expansion solutions are proposed by the QA approaches: use
of synonyms or taxonomy's members, stemming or use of
derivatives\dots

Dubois' dictionary contains some synonyms linked with a sense of
the word they are synonymous with. But these synonyms are too few
to provide sufficient enrichments. The system needs one or more
synonyms dictionaries to complete Dubois' gaps. No synonyms
dictionary shares out the synonyms by sense of the entry, except
\textit{EuroWordNet} \cite{Catherin99}. But \textit{EuroWordNet}'s
sharing out into senses does not match Dubois' senses. Thus the
question is to distribute the available synonyms of each word to
the right sense in Dubois'.

The stemming, which considers two words with the same stem nearly
synonyms, is too unpredictable to be used in a methodology that
tries to avoid noise.
As Dubois' provides instructions to form derivatives from lemma
and suffixes for some senses, the derivation is preferred to the
stemming. But the instructions are often vague, and indicate only
the suffix to use and the new part-of-speech. It is not sufficient
to be used automatically. Thus the derivation procedure needs an
extra tool able to propose derivatives, including the right one.
Dubois' information is sufficient to filter and classify them.

Finally, Dubois' does not provide taxonomy, and the French
resources containing a semantic hierarchy do not supply contextual
information. The taxonomy has to be found in another resource,
which is not consistent with the reference dictionary. The
compatibility between senses of all these resources is the
objective.

\section{How to make the dictionaries compatible}

The main difficulty is to share out information collected from
extra dictionaries. The dictionaries are incompatible with
Dubois', but new data have to be distributed following the senses
of the entries of the reference dictionary.

\subsection{Synonyms}

%

Three resources are at my disposal: Bailly's dictionary
\cite{Bailly47}, an electronic dictionary designed by Memodata,
and the French \textit{EuroWordNet} \cite{Catherin99}. The
expansion methods commonly use all the available synonyms for a
word, but my approach has to keep only the synonyms corresponding
to the current sense of the word. For each considered sense for a
word, Dubois' provides semantic features: a semantic class and an
application domain.

The synonyms from the extra dictionaries are proposals. A proposal
for a lemma in Dubois' dictionary is kept for a given sense only
if one sense at least of the Dubois' entry corresponding to the
proposal matches the semantic features of the given sense.
If no sense of the proposal matches the semantic features of the
given sense, the proposal is rejected for this sense.

\begin{figure}[htb]
\begin{center}\small{
\mbox{\parbox{7cm}{
\begin{tabular}{lll}
\multicolumn{3}{l}{\textsf{Semantic features of the entry:}}\\
& Domain & Class \\
ravir (2) & SOC & S4 \\
&&\\
\multicolumn{3}{l}{\textsf{Semantic features of synonyms:}} \\
& Domain & Class \\
charmer & PSY & P2 \\
\textbf{voler} & \textbf{SOC} & \textbf{S4} \\
\end{tabular}
}}} \caption{Selection of the synonyms.} \label{exSyno}
\end{center}
\end{figure}

In figure \ref{exSyno}, the problem is to determine which proposal
matches the word \textit{ravir} in the sense nb 2 \textit{voler}
(to steal). The semantic features of this sense are the
application domain \texttt{SOC} (sociology) and the semantic class
\texttt{S4} (to grip, to own). The proposal \textit{charmer},
which features are \texttt{PSY} (psychology) and \texttt{P2}
(psychological verb) does not match the features of \textit{ravir}
2. The proposal \textit{d\'{e}rober} in its second and fourth
senses has the same features. This proposal is confirmed for
\textit{ravir} in sense nb 2. It will be used as enrichment when
the sense nb 2 of \textit{ravir} is detected in an utterance by
the semantic disambiguator. This procedure is applied for all the
proposed synonyms for all the senses of each entry in Dubois'.

\subsection{Derivatives}

The derivation field in the Dubois' provides sufficient
indications to recognize the stipulated derivatives of an entry in
a determined sense. Thus, the need is a resource or a tool
providing all the potential derivative from a word. Resources are
rare and incomplete for French, but I have to my disposal a tool
\cite{GaussierAl00} able to construct suffixal derivatives from a
word.
If the only constraint requires the derivatives belong to the
lexicon, all the right suffixal forms are provided among the
incorrect proposals. When all the proposals are produced, the
suffix of each proposal is compared with the instructions in the
dictionary. When they match, the derivative is kept for the
current sense. If not, the derivative is rejected.

\begin{figure}[htb]
\begin{center}\small{
\mbox{\parbox{7cm}{
\begin{tabular}{l|c|l}
\multicolumn{3}{l}{\textsf{Derivatives for the verb \textit{couper}:}}\\
\textsf{Proposed} & \textsf{Instruction} & \textsf{Retained} \\
\textsf{derivatives} & \textsf{sense nb 1} & \textsf{derivatives}\\
\hline
coupure & \textit{ure} & \textbf{coupure} \\
coupable & --- & removed \\
coupage & (\textit{age} sense 5) & removed \\
coupeur & \textit{eur} & \textbf{coupeur} \\
coupant & \textit{ant} & \textbf{coupant} \\
\dots & \dots & \dots \\
\end{tabular}
}}} \caption{Selection of the derivatives.} \label{exDeriv}
\end{center}
\end{figure}

The figure \ref{exDeriv} shows the working of the method. For the
verb \textit{couper} in the sense \textit{trancher} (to cut, to
slice), Dubois' indicates derivatives with suffixes \textit{-ure}:
\textit{coupure} (break), \textit{-ant}: \textit{coupant} (sharp)
and \textit{-eur}: \textit{coupeur} (cutter). But no instruction
is given for a suffix \textit{-able}. The wrong derivative
\textit{coupable} (guilty) is rejected.

\subsection{Taxonomy}

Only two resources containing taxonomy exist for French.
\textit{AlethDic} \cite{gsi93} is known for its very bad quality.
The hierarchy is neither very deep, nor very large. The semantic
relations are not strictly defined inside the hierarchy. Because
of this, I rejected \textit{AlethDic}.

The other resource is \textit{EuroWordNet}. Two kind of taxonomic
relations are defined: hyperonymy (and hyponymy), and meronymy
(and holonymy). The other semantic relations of this resource fall
outside the scope of this paper.

The taxonomic relations link synsets together. The synsets contain
synonymous words for at least one of their senses. The taxonomy is
usable by the QA system only if the sense of the whole synset can
be identified, and if the sense matches at least one of the sense
of the word under consideration in Dubois' dictionary.

So each word in Dubois' has to be linked with a synset to be
inserted into a taxonomic hierarchy. That amounts to match senses
in Dubois' and synsets in \textit{EuroWordNet}. We already have
some senses in Dubois' matching sets of synonyms in
\textit{EuroWordNet}. It is easy to use the additional synonyms
from \textit{EuroWordNet} to set up a correspondence between sense
of Dubois' dictionary and synsets of \textit{EuroWordNet}.

The procedure is to examine all the synsets where a considered
word appears. For each of its sense, if the majority of the
synonyms obtained from \textit{EuroWordNet} are contained in a
synset, the meaning illustrated by the synset and this sense of
the word are considered to be equivalent. In this case, the word
under consideration is inserted in this place into the taxonomic
hierarchy. Otherwise, the synset is not seen to match the sense,
and it is rejected.

\section{Experienced problems}

The difficulties met differ for each kind of processed data. In
the sharing out of the synonyms, the system cannot determine
automatically the meaning of a multiword expression. Dubois' only
lists single words, and no semantic feature can be allocated to a
multiword expression. A multiword proposal is considered to have
all the meaning of the word to which it is synonym.

I have no real evaluation of this procedure: the division into
senses of the reference dictionary is as always open to doubt.
Considering the result, examiners never agree with each synonyms
for a sense. But when they agree (three examiners where
consulted), they where satisfied by more than 80\% of the
synonyms.

The derivation tool provides nearly all the derivatives from a
word when no constraint is defined. Most of the wrong derivatives
(about 97\%) are screened by the instructions supplied by Dubois'
dictionary. However, these figure are not valid for short words:
the tool is designated in such a way that derivatives with a
radical shorter than 3 letters are generally wrong.
Moreover, the instructions are often incomplete in the dictionary,
above all nominal entries.

The promising procedure using taxonomy, presented above, is still
a suggestion. I am facing with the problem that
\textit{EuroWordNet} covers only a small part of the French
lexicon. A more proper trial should use \textit{WordNet}
\cite{FellbaumA98}, that covers a huge part of the English
lexicon, in an English QA application.


\section{Conclusion}

In this paper, I present an original method to merge several
dictionaries in such a way that all the informative fields become
semantically consistent. This need comes from an expansion method
for QA, which uses as enrichment only the synonyms, derivatives,
and taxonomy that match the sense determined by a semantic
disambiguator. The method to share out information is particular
for each kind of enrichment.
An analysis of the experienced problems shows the interest of the
method, and insists on the points that still have to be tackled.

In a more general perspective, it is known that no perfect
dictionary exists. Each dictionary used by this method has gaps.
The method used to mix information is filtering data from extra
dictionaries by data from a reference dictionary, but errors in
the reference are passed on to the added data. The right solution
should be to use more than one reference dictionary.

%

%
%

\bibliographystyle{acl}
\bibliography{biblio}

\end{document}